\documentclass[twocolumn,prb,showpacs]{revtex4}
\usepackage{amsmath}
\usepackage{amssymb}
\usepackage{graphicx}

\input{tcilatex}

\begin{document}

\title{Effects of spin imbalance on the electric-field driven quantum
dissipationless spin current in $p$-doped Semiconductors}
\author{Liangbin Hu$^{1,2}$, Ju Gao$^1$, and Shun-Qing Shen$^1$ }
\affiliation{$^1$Department of Physics, The University of Hong Kong, Pokfulam Road, Hong
Kong, P.R.China \\
$^2$Department of Physics, South China Normal University, Guangdong 510631,
P.R.China}
\date{October 18, 2004}
\begin{abstract}
It was proposed recently by Murakami et al. [Science \textbf{301},
1348(2003)] that in a large class of $p$-doped semiconductors, an applied
electric field can drive a quantum dissipationless spin current in the
direction perpendicular to the electric field. In this paper we investigate
the effects of spin imbalance on this intrinsic $spin$ Hall effect. We show
that in a real sample with boundaries, due to the presence of spin imbalance
near the edges of the sample, the spin Hall conductivity is not a constant
but a sensitively $position$-$dependent$ quantity, and due to this fact, in
order to take the effects of spin imbalance properly into account, a
microscopic calculation of both the quantum dissipationless spin\ Hall
current and the spin accumulation on an equal footing is thus required.
Based on such a microscopic calculation, a detailed discussion of the
effects of spin imbalance on the intrinsic spin Hall effect in thin slabs of 
$p$-doped semiconductors are presented.
\end{abstract}

\pacs{73.43.-f, 72.25.Dc, 72.25.Hg, 85.75.-d}
\maketitle


\section{Introduction}

Efficient injection and coherent control of spins in non-magnetic
semiconductors represent two principal challenges in the emerging field of
spintronics, a new paradigm of semiconductor electronics based on the
utilization of the electron's spin degree of freedom\cite{Prinz98}. At a
first glance, it seems a trivial thing to inject spins into non-magnetic
semiconductors by use of ferromagnetic metals as sources. However, in
reality it is not practical because most of the spin polarizations will be
lost at the interface between metal and semiconductor due to the large
conductivity mismatch\cite{Hammar99,Schmidt00}. A possible approach that can
solve this problem is to replace ferromagnetic metals by ferromagnetic
semiconductors ( such as Ga$_{1-x}$Mn$_{x}$As ) as sources of spin injection%
\cite{Ohno99,Fiederling99,Mattana03}, but for practical use at room
temperature, the Curie temperatures of ferromagnetic semiconductors are
still too low. Due to such difficulties, how to achieve efficient injection
of spins into non-magnetic semiconductors at room temperature remains an
open question and more great efforts are still needed. Recently, based on
the Luttinger effective Hamiltonian\cite{Luttinger}, Murakami et al.
theoretically predicted that an extraordinary $spin$ Hall effect may occur
in a large class of $p$-doped semiconductors ( such as Si, Ge, and GaAs ),
which means that in such a semiconductor, an applied electric field can
drive a substantial amount of quantum dissipationless spin current in the
direction perpendicular to the electric field, and the spin current does not
decrease substantially even at room temperature\cite{Murakami03, Murakami2}.
This effect might reveal a new way for achieving efficient spin injection in
non-magnetic semiconductors at room temperature without the need of
ferromagnetic metals and may also find some other important applications in
spintronics. Prior to the discovery of this effect, a similar effect was
also predicted by Hirsch\cite{Hirsch99} and discussed extensively by several
other authors\cite{Zhang00, Hu03}. From the theoretical points of view, the
effect conceived by Hirsch is an $extrinsic$ spin Hall effect, which is
caused by the spin-orbit dependent anisotropic scatterings from impurities
but not an intrinsic property of a material, and it will disappear
completely in the absence of impurity scatterings. The spin current
generated by the extrinsic spin Hall effect was shown to be rather small\cite%
{Hirsch99, Zhang00, Hu03}, so it is of little use in the problem of spin
injections in non-magnetic semiconductors. Unlike the extrinsic spin Hall
effect, the spin Hall effect proposed in Refs.[8-9] is purely $intrinsic$,
which arises from the intrinsic spin-orbit coupling in the valence bands of $%
p$-doped semiconductors and does not rely on any spin-orbit dependent
anisotropic scatterings from impurities. From a more profound point of view,
this effect has a deep topological character and shares some basic features
with the quantum Hall edge current both physically and mathematically\cite%
{Murakami03, Murakami2}. For example, just like the case of quantum Hall
effect\cite{Thouless, Hatsugai, Niu}, the spin Hall conductivity due to this
effect is a dissipationless transport coefficient and can be expressed as an
integral over all states below the Fermi energy, and the contribution of
each state can be expressed entirely in terms of the curvature of a gauge
field in momentum space\cite{Murakami03, Murakami2}. Due to such features,
the spin current generated by this intrinsic spin Hall effect can be very
large ( comparable to the ordinary charge currents) and hence can serve as
an effective source for efficient injections of spins in non-magnetic
semiconductors at room temperature. Very recently, a similar intrinsic spin
Hall effect was also found by Sinova et al. in two-dimensional electron
gases ( 2DEGs) with Rashba spin-orbit coupling\cite{Sinova}. They found that
\ in 2DEGs with Rashba spin-orbit coupling, the dissipationless and
intrinsic spin Hall conductivity will take a universal value as long as both
spin-orbit split bands are occupied. It is anticipated this effect will also
find some important applications in the emerging field of spintronics.

Although some basic concepts about the intrinsic spin Hall effect are clear%
\cite{Murakami03, Murakami2, Sinova}, there are still a number of important
questions which are needed to be further clarified, and in the last year
many theoretical works have been devoted to the study of this extraordinary
effect.\cite{Shen, Schliemann, Rashba, Murakami04, Culcer, Sinitsyn, Jphu,
Inoue, Dimitrova, Burkov} Basically, most of these theoretical works have
been focused on the calculation of the intrinsic spin Hall conductivity. In
the present paper, we present a theoretical investigation on the effects of
spin imbalance on the intrinsic spin Hall effect in $p$-doped
semiconductors. While it was well known both experimentally and
theoretically that in spin-polarized transport phenomena\cite%
{Pratt91,Gijs93, Johnson85, Valet93, Heide} ( including in semiconductor
spintronics devices\cite{Awschalom, Kato,Stephens} ) spin imbalance may have
significant influences on the transports of spins, what influences spin
imbalance will have on the intrinsic spin Hall effect is still a new subject
and has not yet been explored. For the intrinsic spin Hall effect, from both
the experimental and theoretical points of view, a clear understanding of
the effects of spin imbalance would be much desirable because spin imbalance
may not only have some significant influences on the electric-field driven
quantum dissipationless spin current and on its practical applications but
also play a crucial role in the experimental measurement of the effect\cite%
{Murakami03}. In this paper, based on a solid microscopic ground, we will
derive a set of self-consistent spin transport equations which will present
a proper description on the interplay between the spin imbalance and the
electric-field driven quantum dissipationless spin current in the intrinsic
spin Hall effect in $p$-doped semiconductors. Starting from these spin
transport equations and with the help of appropriate boundary conditions,
the quantum dissipationless spin current and the induced nonequilibrium spin
accumulation in an actual sample with boundaries can be calculated
simultaneously on an equal footing. Our results show that the
characteristics of the interplay between the quantum dissipationless spin
current and the spin imbalance in the intrinsic spin Hall effect in $p$%
-doped semiconductors are very different from what was found in usual
spin-polarized transport phenomena ( including in the extrinsic spin Hall
effect ), and some usual concepts about the interplay between spin current
and spin imbalance cannot be applied to the intrinsic spin Hall effect.

The paper is organized as follows: In Sec.II, we will present a microscopic
derivation of the spin transport equations for describing the intrinsic spin
Hall effect in $p$-doped semiconductors. In our derivation, the effects of
spin imbalance will be included explicitly. In Sec.III, by solving these
spin transport equations with the help of appropriate boundary conditions,
the electric-field driven quantum dissipationless spin current and the
induced nonequilibrium spin accumulation in thin slabs of $p$-doped
semiconductors will be calculated explicitly.

\section{Spin transport equations in the presence of spin imbalance}

In a large class of $p$-doped semiconductors such as Si, Ge, and GaAs, the
valence bands are fourfold degenerate at the $\Gamma $ point. In the
momentum representation and taking the hole picture, the valance bands in
such semiconductors can be described by the following Luttinger effective
Hamiltonian\cite{Luttinger, Murakami03}

\begin{equation}
\hat{H}_{0}=\frac{\hbar ^{2}}{2m}[(\gamma _{1}+\frac{5}{2}\gamma _{2})%
\mathbf{k}^{2}-2\gamma _{2}(\mathbf{k}\cdot \mathbf{S})^{2}],
\end{equation}%
where $\mathit{S}_{i}$ is the spin-3/2 matrix, $\gamma _{1}$ and $\gamma _{2%
\text{ }}$are the Luttinger parameters. For a given wave vector $\mathbf{k}$
, the Hamiltonian (1) has two eigenvalues, given by 
\begin{eqnarray}
\epsilon _{H}(\mathbf{k}) &=&\epsilon _{\lambda =\pm 3/2}(\mathbf{k})=\frac{%
\gamma _{1}-2\gamma _{2}}{2m}\hbar ^{2}k^{2}\equiv \frac{\hbar ^{2}k^{2}}{%
2m_{H}}, \\
\epsilon _{L}(\mathbf{k}) &=&\epsilon _{\lambda =\pm 1/2}(\mathbf{k})=\frac{%
\gamma _{1}+2\gamma _{2}}{2m}\hbar ^{2}k^{2}\equiv \frac{\hbar ^{2}k^{2}}{%
2m_{L}},
\end{eqnarray}%
where $\lambda \equiv \hbar ^{-1}\mathbf{k}\cdot \mathbf{S}/k$ is a good
quantum number of the Hamiltonian $\hat{H}_{0}$. The hole bands described by
Eqs.(2-3) are referred to as the light-hole (LH) and heavy-hole (HH) bands,
respectively. When a uniform electric field $\mathbf{E}$ is applied, the
full Hamiltonian will be given by $\hat{H}=\hat{H}_{0}+e\mathbf{E}\cdot 
\mathbf{x}$, where $-e$ is the charge of an electron. The equation of motion
for the light and heavy holes in a uniform electric field has been derived
in much detail in Ref.[8], and in the semiclassical approximation ( i.e.,
the spin is treated as a classical variable and hence commutes with the
current operator ), the following equation of motion was obtained therein, 
\begin{equation}
\dot{k}_{i}=\frac{eE_{i}}{\hbar },\;\dot{x}_{i}=\frac{\hbar k_{i}}{%
m_{\lambda }}+\epsilon _{ijl}\lambda (2\lambda ^{2}-\frac{7}{2})\frac{k_{l}}{%
k^{3}}\dot{k}_{j},
\end{equation}%
where $\epsilon _{ijl}$ is the usual fully antisymmetric tensor in three
dimensions. The occurrence of the last term in Eq.(4) is unusual, it
represents a \textquotedblleft Lorentz force\textquotedblright\ in momentum
space and is a natural generalization of the quantum Hall effect\cite%
{Thouless, Hatsugai, Niu} to three dimensions\cite{Murakami03, Murakami2}.
It is just due to this \textquotedblleft Lorentz force\textquotedblright\ in
momentum space ( which makes the hole velocity noncollinear with its
momentum ) that the applied electric field will drive a quantum
dissipationless spin Hall current in the direction perpendicular to the
electric field. From Eq.(4), one can get that in the low temperature limit
and in the semiclassical approximation, the net spin current due to both the
LH and HH bands will be given by\cite{Murakami03} 
\begin{eqnarray}
J_{j}^{_{i}} &=&\frac{\hbar }{3}\sum_{\lambda ,\mathbf{k}}\dot{x}_{j}\frac{%
\lambda k_{i}}{k}n_{\lambda }(\mathbf{k})  \notag \\
&=&\sigma _{s}^{0}\epsilon _{ijk}E_{k},
\end{eqnarray}%
where $J_{j}^{_{i}}$ denotes the net spin current following to the $x_{j}$
direction with spin parallel to the $x_{i}$ direction, $n_{\lambda }(\mathbf{%
k})$ is the filling of holes in the band with helicity $\lambda $, and $%
\sigma _{s}^{0}$ is the $spin$ Hall conductivity, which is given by 
\begin{equation}
\sigma _{s}^{0}=\frac{e}{12\pi ^{2}}(3k_{H}^{F}-k_{L}^{F}),
\end{equation}%
with $k_{L}^{F}$ and $k_{H}^{F}$ denoting the Fermi wave numbers in the LH
and HH bands, respectively. In obtaining Eqs.(5-6), one has assumed that the
fillings of holes in each band can be described by the simple Fermi-Dirac
equilibrium distribution function. An alternative way of calculating the
intrinsic spin Hall conductivity is by use of the Kubo formula\cite%
{Murakami2, Sinova, Shen, Schliemann, Culcer, Sinitsyn}. Based on the Kubo
formula, as was shown in Ref.[9], the full quantum treatment of the
noncommutativity between the quantum spin and current operator will lead to
a quantum correction to Eq.(6). But if one takes the semiclassical limit,
the result will become the same as was given by Eq.(6).

Eqs.(5-6) are the central results of Refs.[8-9]. They are valid in the
absence of spin imbalance. But in a real sample with boundaries, when a spin
current circulates in it, spin imbalance will be caused inevitably near the
edge of the sample by the spin current, and in the presence of spin
imbalance, the spin current may be significantly different from what was
given by Eqs.(5-6), especially in the regions near the edges of the sample.
The reason for this is that in the presence of spin imbalance, the fillings
of holes in the LH and HH bands may deviate significantly from the
corresponding cases in the equilibrium state. In order to take the effects
of spin imbalance properly into account, one should obtain the distribution
function strictly by solving the Boltzman transport equation, which
describes the changes of the distribution function in a nonequilibrium
state. In a nonequilibrium but steady state, the Boltzman equation reads 
\begin{equation}
\mathbf{\dot{x}}\cdot \nabla f_{\lambda }(\mathbf{x},\mathbf{k})+\mathbf{%
\dot{k}}\cdot \nabla _{\mathbf{k}}f_{\lambda }(\mathbf{x},\mathbf{k}%
)=-\sum_{\lambda ^{^{\prime }}}(\frac{\partial f_{\lambda }}{\partial t}%
)_{\lambda \rightarrow \lambda ^{^{\prime }}}^{(coll.)},
\end{equation}%
where $(\partial f_{\lambda }/\partial t)_{\lambda \rightarrow \lambda
^{^{\prime }}}^{(coll.)}$ is the collision term due to impurity scatterings,
and $\mathbf{\dot{x}}$ and $\mathbf{\dot{k}}$ are the drift velocities of
holes in the real space and in the momentum space, respectively. Similar to
Ref.[8], in this paper we will confine our discussion to the semiclassical
limit and weak external electric field ( i.e., in the linear response regime
) so that the semiclassical equation of motion given by Eq.(4) can be applied%
\cite{Murakami03, Murakami2}. The collision term $(\partial f_{\lambda
}/\partial t)_{\lambda \rightarrow \lambda ^{^{\prime }}}^{(coll.)}$ will be
given by 
\begin{eqnarray}
(\frac{\partial f_{\lambda }}{\partial t})_{\lambda \rightarrow \lambda
^{^{\prime }}}^{(coll.)} & =& -\int \frac{d^{3}\mathbf{k}^{^{\prime }}}{%
(2\pi )^{3}}w_{\lambda ,\lambda ^{^{\prime }}}^{(i)}(\mathbf{k},\mathbf{k}%
^{^{\prime }})\delta (\epsilon _{\lambda }(\mathbf{k})-\epsilon _{\lambda
^{^{\prime }}}(\mathbf{k}^{^{\prime }}))  \notag \\
& \times & \lbrack f_{\lambda }(\mathbf{x},\mathbf{k})-f_{\lambda ^{^{\prime
}}}(\mathbf{x},\mathbf{k}^{^{\prime }})],
\end{eqnarray}%
where $w_{\lambda ,\lambda ^{^{\prime }}}^{(i)}(\mathbf{k},\mathbf{k}%
^{^{\prime }})$ is the probability of a hole to be scattered from the state $%
|\mathbf{k}\lambda \rangle $ into the state $|\mathbf{k}^{^{\prime }}\lambda
^{^{\prime }}\rangle $ due to impurity scatterings, and the impurity
scatterings will be assumed to be isotropic and spin-independent.

In the equilibrium state, the fillings of holes in each band are stable and
can be described by the simple Fermi-Dirac equilibrium distribution
function. When the external electric field is applied and the system turns
into an nonequilibrium but steady state, the fillings of holes in each band
will still be stable but different from what was described by the simple
Fermi-Dirac equilibrium distribution function. In the presence of spin
imbalance, the changes of the fillings of holes in the LH and HH bands will
be caused primarily by two kinds of contributions. The first kind of
contribution is caused by the drifts of holes in the external electric
field, and the second kind of contribution is due to the occurrence of spin
imbalance. In the linear response regime ( i.e., in a weak electric field ),
the deviations of the fillings of holes in each band from the corresponding
cases in the equilibrium state are small, and the two kinds of contributions
will be independent and both be \ proportional to $\partial f^{0}/\partial
\epsilon _{F}$ ( here $f^{0}=\{\exp [\beta (\epsilon _{\lambda }(\mathbf{k}%
)-\epsilon _{F})]+1\}^{-1}$ is the usual Fermi-Dirac equilibrium
distribution function with $\beta $ denoting the inverse of temperature and $%
\epsilon _{F}$ the Fermi level in the equilibrium state ). Considering this
fact and by use of the relaxation time approximation, in the linear response
regime the nonequilibrium distribution function $f_{\lambda }(\mathbf{x},%
\mathbf{k})$ ( in a nonequilibrium but steady state ) can be expressed as
the following, 
\begin{equation}
f_{\lambda }(\mathbf{x},\mathbf{k})=f^{0}+\mu _{\lambda }(\mathbf{x})\frac{%
\partial f^{0}}{\partial \epsilon _{F}}+e\tau \lbrack \mathbf{E}_{\lambda }(%
\mathbf{x})\cdot \mathbf{V}_{\lambda }]\frac{\partial f^{0}}{\partial
\epsilon _{F}},
\end{equation}%
where $\mathbf{V}_{\lambda }=\hbar \mathbf{k/}m_{\lambda }$ is the velocity
of holes; $\tau $ is the total relaxation time of holes due to
impurity-induced random scatterings; $\mathbf{E}_{\lambda }(\mathbf{x})$ is
the total effective field felt by a moving hole in the band with helicity $%
\lambda $, which is the sum of the external electric field $\mathbf{E}$ and
a band-dependent effective field induced by the spin imbalance in the
sample. The detailed definition of $\mathbf{E}_{\lambda }(\mathbf{x})$ and $%
\tau $ will be given below. The second term in Eq.(9) just characterizes the
deviation of the filling of holes in the band with helicity $\lambda $ from
the corresponding case in the equilibrium state due to the occurrence of
spin imbalance in the sample, and the presence of this term is
mathematically equivalent to introducing a band-dependent \textquotedblleft\
shift \textquotedblright\ $\mu _{\lambda }$ in the Fermi level $\epsilon
_{F} $. ( It should be noted that unlike the corresponding cases in usual
spin-polarized transport phenomena, here $\mu _{\lambda }$ does not relate
directly to the spin accumulation because the label $\lambda $ does not
correspond to a fixed spin-polarization direction in real space. ). The
third term in Eq.(9) denotes the change of the filling due to the drifts of
holes in the external electric field and in the presence of impurity
scatterings. By inserting Eq.(9) into Eq.(7) and assuming that the impurity
scatterings are isotropic, the Boltzman equation can be simplified to the
following form, 
\begin{eqnarray}
&&\mathbf{V}_{\lambda }\cdot \lbrack \mathbf{E}+\frac{1}{e}\bigtriangledown
\mu _{\lambda }(\mathbf{x})+\tau \bigtriangledown (\mathbf{E}_{\lambda }(%
\mathbf{x})\cdot \mathbf{V}_{\lambda })]  \notag \\
&=&\sum_{\lambda ^{^{\prime }}(\neq \lambda )}\frac{\mu _{\lambda }(\mathbf{x%
})-\mu _{\lambda ^{^{\prime }}}(\mathbf{x})}{e\tau _{\lambda \lambda
^{^{\prime }}}}+\sum_{\lambda ^{^{\prime }}}\frac{\tau \mathbf{E}_{\lambda }(%
\mathbf{x})\cdot \mathbf{V}_{\lambda }}{\tau _{\lambda \lambda ^{^{\prime }}}%
},
\end{eqnarray}%
where $\tau _{\lambda \lambda ^{^{\prime }}}$ is a characteristic relaxation
time defined by 
\begin{equation}
\tau _{\lambda \lambda ^{^{\prime }}}=[\int \frac{d^{3}\mathbf{k}^{\prime }}{%
(2\pi )^{3}}w_{\lambda ,\lambda ^{^{\prime }}}^{(i)}(\mathbf{k},\mathbf{k}%
^{^{\prime }})\delta (\epsilon _{\lambda }(\mathbf{k})-\epsilon _{\lambda
^{^{\prime }}}(\mathbf{k}^{^{\prime }}))]^{-1},
\end{equation}%
which characterizes the probability ( given by $\tau _{\lambda \lambda
^{^{\prime }}}^{-1}$ ) for a hole in the band with helicity $\lambda $ to be
scattered into the band with helicity $\lambda ^{^{\prime }}$ due to
impurity scatterings. For simplicity, in the following we will assume that
the intra-band-scattering relaxation time $\tau _{\lambda \lambda }\equiv
\tau _{1}$ ( independent of $\lambda $ ) and the inter-band-scattering
relaxation time $\tau _{\lambda \lambda ^{^{\prime }}(\neq \lambda )}\equiv
\tau _{2}$ ( independent of $\lambda $ and $\lambda ^{^{\prime }}$ ), and as
usual, the total scattering probability for a hole ( given by $\tau ^{-1}$,
i.e., the inverse of the total relaxation time of a hole ) can be given by 
\begin{equation}
\tau ^{-1}=\sum_{\lambda ^{^{\prime }}}\tau _{\lambda \lambda ^{^{\prime
}}}^{-1}=\frac{1}{\tau _{1}}+\frac{3}{\tau _{2}},
\end{equation}%
which is assumed to be independent of the band label $\lambda $. Multiplying
both sides of Eq.(10) by $\mathbf{V}_{\lambda }$ and then integrating both
sides with respect to $\mathbf{V}_{\lambda }$ and with the help of Eq.(12),
one can find that the total effective field felt by a moving hole in the
band with helicity $\lambda $ should be given by 
\begin{equation}
\mathbf{E}_{\lambda }(\mathbf{x})=\mathbf{E}+\frac{1}{e}\bigtriangledown \mu
_{\lambda }(\mathbf{x}).
\end{equation}%
Eq.(13) suggests that in the presence of spin imbalance, in addition to the
external electric field $\mathbf{E}$, conduction electrons will also feel an
effective field proportional to the gradient of the band- and
position-dependent shift in the Fermi level. After $\tau $ and $\mathbf{E}%
_{\lambda }(\mathbf{x})$ are determined from Eqs.(12-13), the nonequilibrium
distribution function $f_{\lambda }(\mathbf{x},\mathbf{k})$ will also be
determined by Eq.(9). Then in the semiclassical limit the electric-field
driven quantum dissipationless spin current can be obtained through the
following formula, 
\begin{equation}
J_{j}^{i}(\mathbf{x})=\sum_{\lambda }\int \frac{d^{3}\mathbf{k}}{(2\pi )^{3}}%
[\dot{x}_{j}s_{i,\lambda }(\mathbf{k})]f_{\lambda }(\mathbf{x,k}),
\end{equation}%
where $\dot{x}_{j}=\hbar k_{j}/m_{\lambda }+\epsilon _{jkl}\lambda (2\lambda
^{2}-7/2)k_{l}\dot{k}_{k}/k^{3}$ ( see Eq.(4) ) and $s_{i,\lambda }(\mathbf{k%
})=\hbar (\lambda k_{i}/k)/3$ are the velocity and the spin of a hole with
momentum $\mathbf{k}$ and helicity $\lambda $, respectively. ( Since the
spin-$3/2$ matrix $\mathbf{S}$ in the Hamiltonian (1) is a summation of the
spin angular momentum $\mathbf{s}$ with spin one-half and the atomic orbital
angular momentum $\mathbf{l}$ with spin one, the expectation value of $%
\mathbf{s}$ should be one third of $\mathbf{S}$.\cite{Murakami03,Luttinger}%
). By substituting Eqs.(12-13) into Eq.(9) and then inserting Eq.(9) into
Eq.(14), the following result can be obtained 
\begin{equation}
J_{j}^{i}(\mathbf{x})=\sigma _{s}(\mathbf{x})\epsilon _{ijk}E_{k},
\end{equation}%
where $\sigma _{s}(\mathbf{x})$ is the spin Hall conductivity in the
presence of spin imbalance, which is given by 
\begin{equation}
\sigma _{s}(\mathbf{x})=\sigma _{s}^{0}-\frac{e}{48\pi ^{2}\epsilon _{F}}%
[3k_{H}^{F}\mu _{H}(\mathbf{x})-k_{L}^{F}\mu _{L}(\mathbf{x})].
\end{equation}%
Here $\sigma _{s}^{0}$ is the spin Hall conductivity in the $absence$ of
spin imbalance, which has been defined in Eq.(6), and $\mu _{H}(\mathbf{x}%
)\equiv \mu _{3/2}(\mathbf{x})+\mu _{-3/2}(\mathbf{x})$ and $\mu _{L}(%
\mathbf{x})\equiv \mu _{1/2}(\mathbf{x})+\mu _{-1/2}(\mathbf{x})$.
Eqs.(15-16) show that the effects of spin imbalance on the quantum
dissipationless spin current due to the intrinsic spin Hall effect in $p$%
-doped semiconductors are very different from what was found in usual
spin-polarized phenomena ( including the extrinsic spin Hall effect\cite%
{Hirsch99, Zhang00, Hu03} ). First, in the presence of spin imbalance, the
spin Hall conductivity due to the intrinsic spin Hall effect in $p$-doped
semiconductors might not be a constant but a $position$-$dependent$
quantity. This is a new feature that was not seen before. Second, for the
intrinsic spin Hall effect in $p$-doped semiconductors, the change of the
quantum dissipationless spin current due to the occurrence of spin imbalance
is determined directly by $\mu _{\lambda }(\mathbf{x})$ ( i.e., the
band-dependent \textquotedblleft\ shifts \textquotedblright\ in the Fermi
level ) but is independent of the gradients of $\mu _{\lambda }(\mathbf{x})$%
. This is also significantly different from what was found in usual
spin-polarized transports ( including the extrinsic spin Hall effect). These
unusual characteristics of the intrinsic spin Hall effect in $p$-doped
semiconductors can be understood by the following arguments. According to
Eq.(6), the spin Hall conductivity should be determined uniquely by the
Fermi wave numbers $k_{H}^{F}$ and $k_{L}^{F}$. In the presence of spin
imbalance, because the spin imbalance will induce a position and band
dependent shift $\mu _{\lambda }(\mathbf{x})$ in the Fermi level, the Fermi
wave numbers $k_{H}^{F}$ and $k_{L}^{F}$ will also be position-dependent,
and the changes of $k_{H}^{F}$ and $k_{L}^{F}$ due to the occurrence of spin
imbalance will be determined directly by $\mu _{\lambda }(\mathbf{x})$. Due
to this reason, in the presence of spin imbalance, the spin Hall
conductivity will be a position-dependent quantity, and the change of the
spin Hall current due to the occurrence of spin imbalance will be determined
by $\mu _{\lambda }(\mathbf{x})$ but independent of the gradients of $\mu
_{\lambda }(\mathbf{x})$. Finally, it should be pointed out that because we
have considered only isotropic and spinless impurity scattering, the
mechanism of the generation of the spin Hall current described by
Eqs.(15-16) is still purely intrinsic, though there are some significant
differences between Eq.(16) and Eq.(6). In fact, one can check that in the
linear response regime the impurity scattering term ( i.e., the third term )
in Eq.(9) does not contribute to the spin Hall conductivity given by
Eq.(16). This point will be more clearly seen from the results presented in
Sec.III. Of course, if the impurity scatterings are spin-orbit dependent,
then the total spin current will contain not only the intrinsic part but
also contain an extrinsic part due to the spin-orbit dependent impurity
scatterings through the mechanism proposed by Hirsch\cite{Hirsch99, Zhang00,
Hu03}.

In the ordinary charge Hall effect, the charge Hall current causes $charge$
imbalance in a sample and results in charge accumulation. Similarly, in the
spin Hall effect, the spin imbalance caused by the spin Hall current will
result in nonequilibrium spin accumulation in a sample. Corresponding to the
quantum spin Hall current given by Eqs.(15-16), the nonequilibrium spin
accumulation induced by the quantum spin Hall current can be obtained as the
following, 
\begin{eqnarray}
S_{i}(\mathbf{x}) &=&\sum_{\lambda }\int \frac{d^{3}\mathbf{k}}{(2\pi )^{3}}%
s_{i,\lambda }(\mathbf{k})f_{\lambda }(\mathbf{x,k})  \notag \\
&=&\epsilon _{ijl}\frac{E_{l}\hbar ^{2}}{16e\epsilon _{F}^{2}}\frac{\partial 
}{\partial x_{j}}[C_{L}\mu _{L}(\mathbf{x})-3C_{H}\mu _{H}(\mathbf{x})],
\end{eqnarray}%
where $C_{L}=e^{2}\tau (k_{L}^{F})^{3}/6\pi ^{2}m_{L}$ and $C_{H}=e^{2}\tau
(k_{H}^{F})^{3}/6\pi ^{2}m_{H}$ are the ordinary charge conductivities of
the light holes and the heavy holes, respectively. Eqs.(15-17) show that in
the intrinsic spin Hall effect in $p$-doped semiconductors, both the quantum
dissipationless spin current and the spin accumulation will depend
sensitively on $\mu _{\lambda }(\mathbf{x})$, i.e., the band-dependent
\textquotedblleft\ shifts \textquotedblright\ in the Fermi level. To find
out the equations that $\mu _{\lambda }(\mathbf{x})$ should satisfy, one can
substitute Eqs.(12-13) into Eq.(10) and integrate both sides of Eq.(10) with
respect to $\mathbf{k}$, then one will arrive at the following equation 
\begin{equation}
\bigtriangledown ^{2}\mu _{\lambda }(\mathbf{x})=\frac{1}{D_{\lambda }^{2}}%
[4\mu _{\lambda }(\mathbf{x})-\mu _{H}(\mathbf{x})-\mu _{L}(\mathbf{x})],
\end{equation}%
where $D_{\lambda }=V_{\lambda }^{F}\sqrt{\tau _{2}\tau /3}$ is a
characteristic hole diffusion length and $V_{\lambda }^{F}$ is the
band-dependent Fermi velocity. In addition to Eq.(18), $\mu _{\lambda }(%
\mathbf{x})$ should also satisfy the charge neutrality condition, which
requires that the net changes of the charge density due to the flow of the
quantum spin current, given by $\delta \rho _{c}=-e\sum_{\lambda }\int \frac{%
d^{3}\mathbf{k}}{(2\pi )^{3}}[f_{\lambda }(\mathbf{x,k})-f^{0}(\epsilon
_{\lambda }(\mathbf{k}))]$, should be zero. This leads to the following
equation 
\begin{equation}
\mu _{H}(\mathbf{x})=-(\frac{m_{L}}{m_{H}})^{3/2}\mu _{L}(\mathbf{x}).
\end{equation}%
Eqs.(15)-(19) are the central results of the present paper. They constitute
a set of self-consistent equations from which both the quantum
dissipationless spin current and the spin accumulation due to the intrinsic
spin Hall effect in a real sample of $p$-doped semiconductors with
boundaries can obtained simultaneously with the help of appropriate boundary
conditions.

\section{Intrinsic spin Hall effect in thin slabs of $p$-doped semiconductors%
}

Eqs.(15-19) are rather general and in principle they can be applied to
samples with any kind of geometries. In the experimental measurement of the
Hall effect ( including the spin Hall effect ), a thin slab geometry ( i.e.,
the $Hall$ bar ) is usually applied. In this section, starting from
Eqs.(15-19), we will present a detailed theoretical investigation on the
intrinsic spin Hall effect in a thin slab of $p$-doped semiconductors. We
assume that the longitudinal direction of the slab is along the $z$ axis and
the transverse direction along the $y$ axis and the normal of the surface
along the $x$ axis, respectively, and an external electric field $E_{z}$ is
applied in the longitudinal direction of the slab. The thickness of the slab
is assumed to be much smaller than the hole diffusion length $D_{\lambda }$
and the length of the slab is assumed to be much larger than the width, so
that only spin current flowing to the $y$ direction ( i.e., in the
transverse direction of the slab ) with spin parallel to the $x$ direction
need to be considered. The two boundaries of the slab are assumed to be
located at $y=\pm w/2$, and $w$ is the width of the slab. In general, it is
very difficult to solve Eqs.(15-19) analytically. In order to get some
explicit expressions for the spin Hall current and the spin accumulation, we
assume that in Eqs.(15-19) the hole diffusion length $D_{\lambda }$ is $%
\lambda $-independent ( i.e., $D_{\lambda }\equiv D$ ) and $m_{L}\simeq m_{H}
$. Then from Eqs.(18-19) one can see that $\mu _{H}(y)\equiv \mu
_{3/2}(y)+\mu _{-3/2}(y)$ and $\mu _{L}(y)\equiv \mu _{1/2}(y))+\mu
_{-1/2}(y)$ can be expressed as 
\begin{equation}
\mu _{H}(y)\simeq -\mu _{L}(y)=Ae^{y/2D}+Be^{-y/2D},
\end{equation}%
where $A$\ and $B$ are two constant coefficients that need to be determined
by the appropriate boundary condition. In this paper, we will consider the
transverse open circuit boundary condition. In the transverse open circuit
boundary condition, the spin Hall current will be zero at the two boundaries
of the slab, i.e., $J_{y}^{x}(y=\pm w/2)=0$. Substituting Eq.(20) into
Eqs.(15-16), the spin Hall current $J_{y}^{x}(y)$ can be expressed as a
function of the coefficients $A$ and $B$. Then by use of the transverse open
circuit boundary condition, the coefficients $A$ and $B$ can be determined,
and one can get that 
\begin{equation}
A=B=\frac{2\epsilon _{F}(3k_{H}^{F}-k_{L}^{F})}{3k_{H}^{F}+k_{L}^{F}}\frac{1%
}{\cosh (w/2D)}.
\end{equation}%
After the coefficients $A$ and $B$ are determined, the spin Hall current $%
J_{y}^{x}(y)$ and the spin Hall conductivity $\sigma _{s}(y)$ will also be
obtained by inserting Eq.(20) into Eqs.(15-16), and the results are given by 
\begin{eqnarray}
J_{y}^{x}(y) &=&\sigma _{s}(y)E_{z}, \\
\sigma _{s}(y) &=&\sigma _{s}^{0}[1-\frac{\cosh (y/2D)}{\cosh (w/4D)}].
\end{eqnarray}%
Eqs.(22-23) show that, in the presence of spin imbalance, both the spin Hall
current and the spin Hall conductivity might be highly position-dependent
and might also depend sensitively on the hole diffusion length $D$ and the
width $w$ of the sample. The spin Hall conductivity $\sigma _{s}(y)$ will be
maximum at the center of the sample ( i.e., at $y=0$ ) and tend to be zero
at the edges of the sample. Two limiting cases will be especially
interesting. The first case is that the hole diffusion length $D$ is much
larger than the width $w$ of the sample. In this limiting case the spin Hall
current will be very small, i.e., $\sigma _{s}(y)\simeq 0$ everywhere. The
second interesting case is that $w\gg D$. In this limiting case, the maximum
value of the spin Hall conductivity will be given by $\sigma _{s}(y)\simeq
\sigma _{s}^{0}$ ( at $y=0$ ) and $\sigma _{s}(y)\rightarrow 0$ as $%
y\rightarrow \pm w/2$. These features can be seen clearly from Fig.1, where
we have plotted the position dependence of the spin Hall conductivity $%
\sigma _{s}(y)$ in three cases with different ratios of $w/D$. \ From Fig.1
and Eq.(23), one can see clearly that if no boundaries exist ( i.e., $%
w\rightarrow \infty $ and hence no spin imbalance occurs ), the spin Hall
conductivity will be a constant and return to the same result as was given
by Eq.(6), i.e., the spin Hall conductivity will not be changed by weak
isotropic and spinless impurity scatterings. This is in agreement with
Ref.[8] and also in agreement with the result obtained by a more accurate
calculation performed in Ref.[20]. It is interesting to note that recently a
similar conclusion was also obtained for the intrinsic spin Hall effect in
2DEGs with Rashba spin-orbit coupling by both numerical simulations and
analytical calculations, which suggest that in the presence of weak (
isotropic and spinless ) impurity scatterings, the intrinsic spin Hall
conductivity in a Rashba two-dimensional electron gas should still take a
universal value, proving that the sample size exceeds the localization length%
\cite{Dimitrova, Burkov}. Of course, it should be pointed out that at
present different views also exist on this problem. For example, in Ref.[24]
it was argued that the spin-orbit-coupling induced intrinsic spin Hall
current in a Rashba two-dimensional electron gas should vanish in the
presence of impurity scatterings, even if the impurity scatterings are weak
and spinless.

\begin{figure}[tbp]
\includegraphics[width=8.5cm]{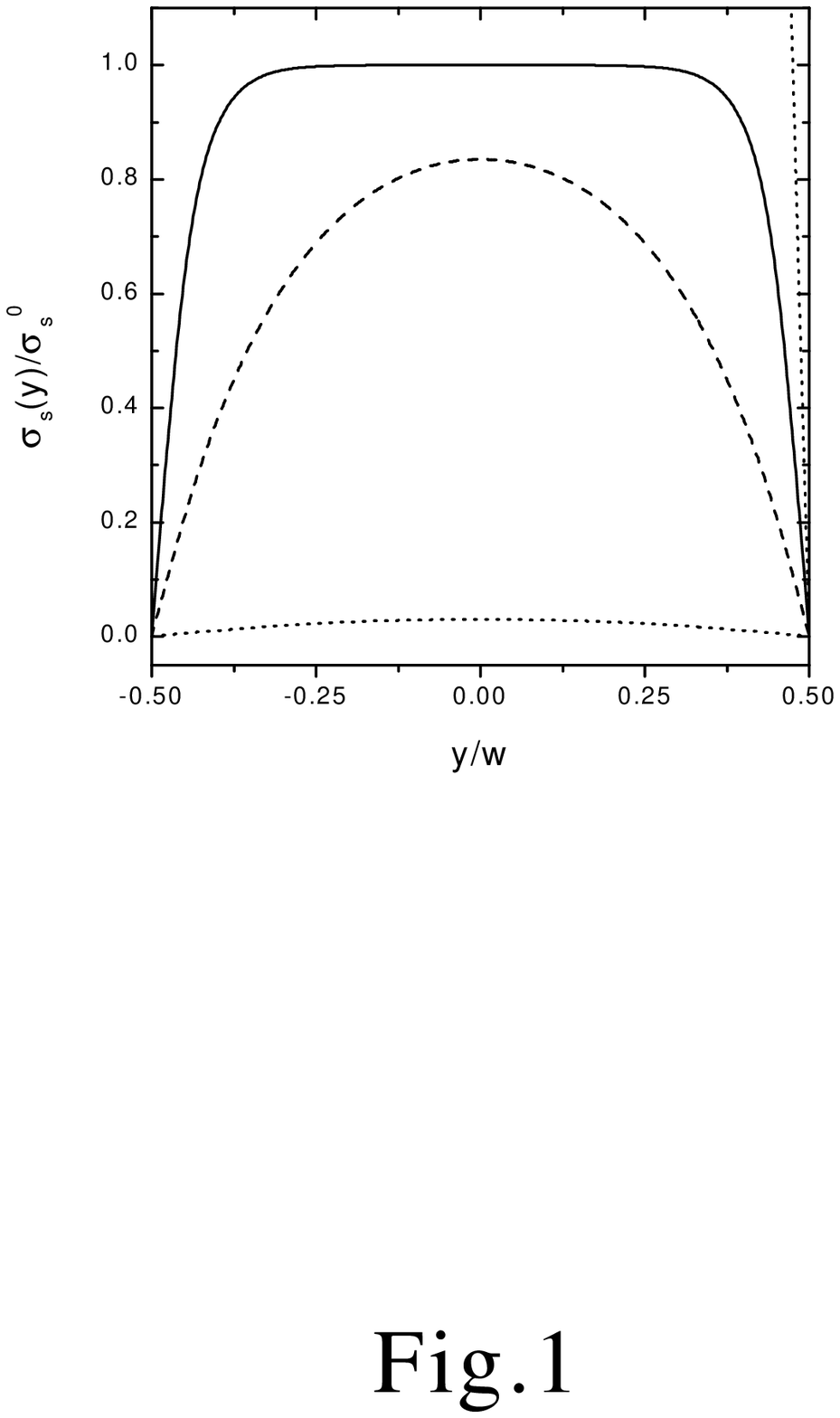}
\caption{
Illustration of the position dependences of the spin Hall
conductivity $\sigma _s(y)$ in three cases with different ratios of $w/D$. ( 
$w/D=50$ for the solid line, $w/D=10$ for the dashed line, and $w/D=1$ for
the dotted line. $\sigma _s(y)$ is normalized by $\sigma _s^0$, i.e., the
spin Hall conductivity in the absence of spin imbalance.)
}
\end{figure}

The quantum dissipationless spin current generated by the intrinsic spin
Hall effect does not carry charges ( i.e., it is a $pure$ spin current), so
it is very difficult to measure the quantum dissipationless spin current
directly. An indirect but much more convenient way to detect the quantum
dissipationless spin current is to measure the nonequilibrium spin
accumulation induced by the quantum dissipationless spin current. The
nonequilibrium spin accumulation induced by the quantum dissipationless spin
Hall current in a thin slab of $p$-doped semiconductors can be got by
inserting Eqs.(20-21) into Eq.(17), and the following result can be
obtained, 
\begin{equation}
S_{x}(y)=\frac{3\pi ^{2}\hbar ^{2}\sigma _{s}^{0}E_{z}(C_{L}+3C_{H})\sinh
(y/2D)}{e^{2}\epsilon _{F}(k_{L}^{F}+3k_{H}^{F})D\cosh (w/4D)}.
\end{equation}%
Eq.(24) shows that the spin accumulation will be linearly proportional to
the spin Hall conductivity $\sigma _{s}^{0}$ and also depend sensitively on
the ordinary charge conductivities $C_{L}$ and $C_{H}$ of the light and
heavy holes. It also have a sensitive dependence on the hole diffusion
length $D$ and the sample width $w$. \ According to Eq. (24), for a
infinitely large sample without boundaries ( i.e., $w\rightarrow \infty $ ),
no spin accumulation will appear ( i.e., $S_{x}(y)=0$ for any finite $y$ ).
This is different from what was found in a Rashba two-dimensional electron
gas, \ where it was found that the application of an in-plane electric field
would induce a homogeneous nonequilibrium spin accumulation without resort
to the boundary effects. \cite{Edelstein, Inoue2, Add} \ From Eq.(24), one
can see that for an actual sample with boundaries, the spatial distribution
of the spin accumulation due to the intrinsic spin Hall effect would be
highly inhomogeneous. The spin accumulation will be maximum at the edges of
the sample and vanish near the center of the slab, and the spin accumulation
at the edges of the sample will increase with the increase of the sample
width $w$. This has been illustrated in Fig.2. From Fig.2 one can see that
if the sample width $w$ is much smaller than the hole diffusion length $D$,
the spin accumulation induced by the quantum spin Hall current will be very
small. On the other hand, if the sample width $w$ is much larger than the
hole diffusion length $D$, the spin accumulation at the edges of the sample
will be almost a constant, independent of the sample width. This will be a
merit for the experimental measurement of the intrinsic spin Hall effect. To
obtain a quantitative estimation on the order of the magnitude of the spin
accumulation induced by the quantum dissipationless spin Hall current in a
real sample, let us consider some actual experimental parameters. We take
the ordinary conductivity $C_{L,H}\sim 10^{2}\Omega ^{-1}$cm$^{-1}$ and the
hole diffusion length $D\sim 10$nm and $\hbar /\epsilon _{F}\sim 1$f$\sec $.
These parameters are typical of the holes in GaAs with the hole density $%
n\sim 10^{19}$cm$^{-3}$. The width $w$ of the sample is assumed to be $%
100\mu $m ( much larger than the hole diffusion length ) and a current
density $j_{x}\sim 10^{4}$A$/$cm$^{2}$. By use of the parameters listed
above, from Eq.(24) it can be estimated that the spin accumulation at the
edges of the sample will be on the order of $10^{13}-10^{15}\mu _{B}$cm$%
^{-2} $. Such magnitudes should be large enough to be measured by some
ordinary experimental methods, for example, by the method proposed in
Refs.[8-11]. Finally, it should be pointed out that a rough estimation of
the spin accumulation due to the quantum dissipationless spin Hall current
was also presented in the supporting online material for Ref.[8] based on a
simple analysis by use of the usual spin diffusion equation, but there are
some significant differences between the results obtained in the present
paper and the corresponding results reported therein. This can be seen by
making a comparison between Eq.(24)\ obtained in the present paper and
Eq.(S16) presented in the supporting online material for Ref.[8]. For
example, according to Eq.(24) obtained in the present paper, the spin
accumulation will not only depend on the spin Hall conductivity but also
depend sensitively on the ordinary charge conductivities of the light and
heavy holes; however, according to Eq.(S16) in the supporting online
material for Ref.[8], the spin accumulation will only depend on the spin
Hall conductivity but is independent of the ordinary charge conductivities
of the light and heavy holes. Our results show that though the mechanism of
the intrinsic spin Hall effect is purely intrinsic, impurity scatterings
might have some significant influences on the effect in a real sample with
boundaries, and the usual spin diffusion equation might not be very suitable
for describing this effect. In fact, from the microscopic calculation
presented in Sec.II, one can see that in general the quantum dissipationless
spin Hall current ( given by Eqs.(15-16) ) and the spin accumulation ( given
by Eq.(17) ) due to the intrinsic spin Hall effect do not satisfy the usual
spin diffusion equation.

\begin{figure}[tbp]
\includegraphics[width=8.5cm]{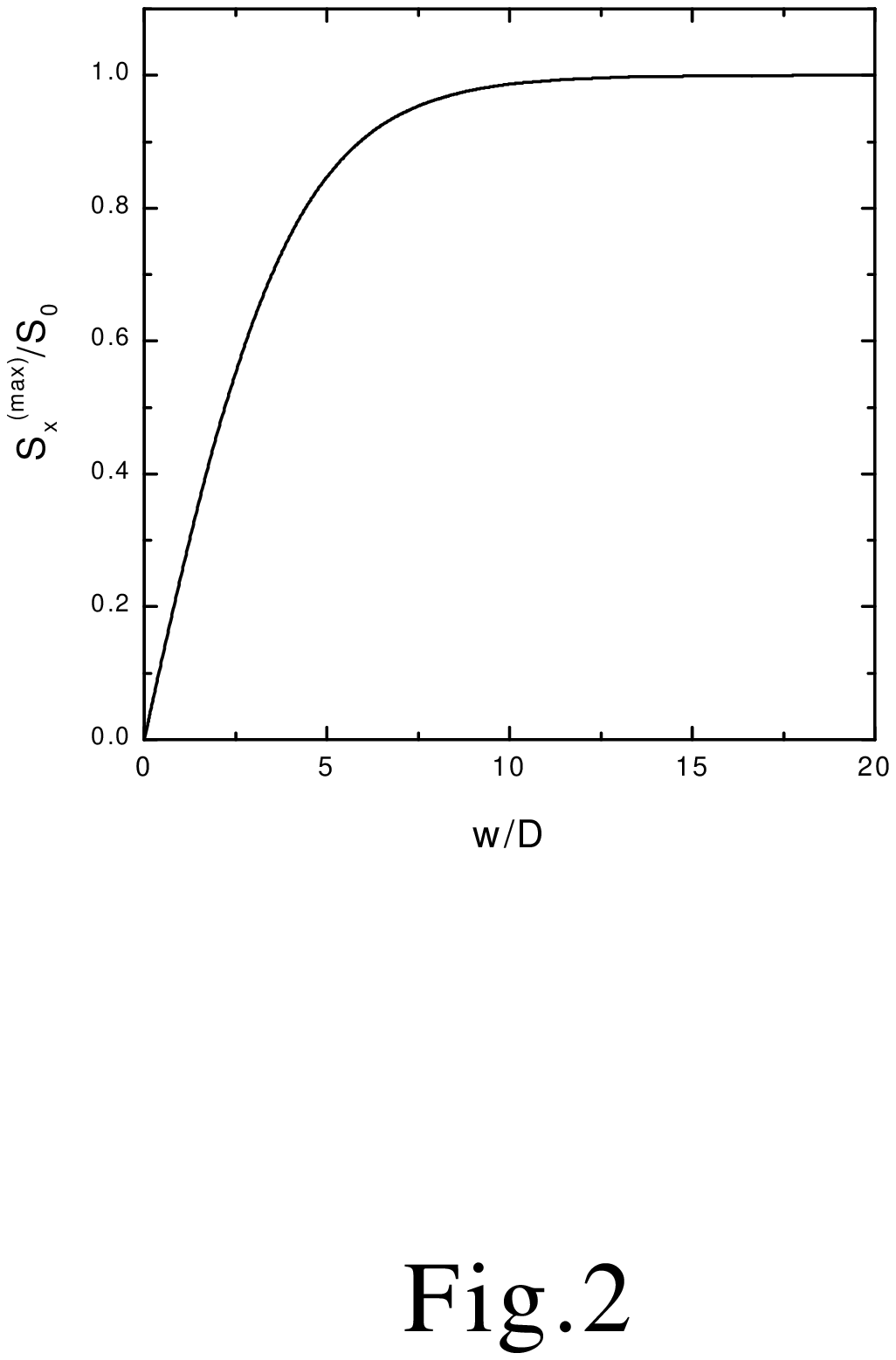}
\caption{
Illustration of the changes of the spin accumulation at the edges of
a sample with the variation of the sample width. ( The spin accumulation is
normalized by $S_0=\frac{3\pi ^2\hbar ^2\sigma _s^0E_z(C_L+3C_H)}{%
e^2\epsilon _FD(k_L^F+3k_H^F)}$. ) 
}
\end{figure}

In conclusion, in this paper we have presented a detailed theoretical
investigation on the effects of spin imbalance on the intrinsic spin Hall
effect in $p$-doped semiconductors. We have shown that in a real sample with
boundaries, the spin Hall conductivity might not be a constant but a
sensitively position-dependent quantity due to the occurrence of spin
imbalance near the edges of the sample, and in order to take the effects of
spin imbalance properly into account, a microscopic calculation of both the
quantum dissipationless spin current and the spin accumulation based on an
equal footing is thus required. We stress that some usual concepts about the
interplay between spin current and spin imbalance might not be suitable for
describing the intrinsic spin Hall effect. After some modifications, the
theory presented in this paper might also be applied to investigate the
effects of spin imbalance in the intrinsic spin Hall effect in 2DEGs with
Rashba spin-orbit coupling. Finally, it should be pointed out that though in
the last year many theoretical works have been devoted to the study of the
intrinsic spin Hall effect, many controversial issues still exist concerning
some fundamental aspects of this extraordinary effect. Among them, a big
controversial issue is that what is the correct definition of spin current
in materials with intrinsic spin-orbit coupling\cite{Murakami2, Sinova,
Rashba, Culcer}. As was argued in Ref.[9] and in Ref.[19], there are some
difficulties with the conventional definition of spin current in
spin-orbit-coupled systems, but it seemed that up to now there are still no
unanimous views about this question\cite{Murakami2, Sinova, Shen, Rashba,
Schliemann, Culcer, Jphu}. ( In the present paper we have used the same
definition of Refs.[8] ). Because no unambiguous experimental detections
have ever been done, on the present stage such controversial issues are
difficult to be clarified unambiguously by pure theoretical arguments. But
it could be anticipated that by combining future experimental results with
more accurate theoretical investigations, these controversial issues should
be able to be clarified unambiguously in the near future.

\bigskip

\textbf{Acknowledgements}

S. Q. Shen is supported by a grant from the Research Grant Council of Hong
Kong. L. B. Hu is supported in part by the National Science Foundation of
China (Grant No.10474022).


\begin{thebibliography}{99}
\bibitem{Add} When an in-plane electric field is applied in \ a
two-dimensional electron gas with Rashba spin-orbit coupling, the spin-orbit
coupling itself ( without resort to the boundary effects ) will induce a
homogeneous spin accumulation with the spin-polarization direction
perpendicular to the electric field. This effect was called the kinetic
magnetoelectric effect\cite{Inoue2, Edelstein}. This effect should not be
confused with the usual spin accumulation phenomena widely studied in the
literatures. Form theoretical viewpoints, the usual spin accumulation
phenomena are caused by the boundary effects\cite{Pratt91, Gijs93,
Johnson85, Valet93, Heide, Awschalom}, but the kinetic magnetoelectric
effect is caused by the combined action of the spin-orbit coupling and the
electric field and has no relation with the boundary effects.
\end{thebibliography}
\end{document}